\documentclass{epl}
\usepackage{amssymb}

\title{Scale-Free Dynamics Emerging from Information Transfer}
\shorttitle{Criticality Emerging from Information Transfer}

\author{M.Y. Choi\inst{1,2} \and Beom Jun Kim\inst{3} \and B.-G.
   Yoon\inst{4} \and Hyunggyu Park\inst{2}}
\shortauthor{M.Y.Choi {\it et al.}}
\institute{
   \inst{1} Department of Physics, Seoul National University, Seoul 151-747, Korea \\
   \inst{2} Korea Institute for Advanced Study, Seoul 130-722, Korea \\
   \inst{3} Department of Molecular Science and Technology, Ajou University,
   Suwon 442-749, Korea \\
   \inst{4} Department of Physics, University of Ulsan, Ulsan 680-749, Korea
}

\begin{document}

\maketitle

\begin{abstract}
The dynamics based on information transfer is proposed as an underlying mechanism for
the scale-invariant dynamic critical behavior observed in a variety of systems.
We apply the dynamics to the globally-coupled Ising model, which is analytically 
tractable, and show that dynamic criticality is indeed attained.  
Such emergence of criticality is confirmed numerically in 
the two-dimensional Ising model as well as the globally coupled one
and in a biological evolution model. 
Although criticality is precise only when information transfer is reversible,
it may also be observed even in the irreversible case, 
during the practical time scale shorter than the relaxation time. 
\end{abstract}

There exist rich examples of scale invariance in nature, manifested by
the abundant power-law behaviors of characteristic spectra, 
apparently without fine tuning. This is in sharp contrast with the scale
invariance present exclusively at a phase transition point, 
for which renormalization-group techniques have been developed~\cite{rg}. 
As a possible framework to explain such scale invariance or criticality
without fine tuning, 
the concept of self-organized criticality (SOC) has been proposed and, 
in its most broad sense, used widely~\cite{soc}. 
However, the general theory for ubiquitous emergence of criticality is still lacking 
and most existing works are studies of specific models on the case-by-case basis.
As an attempt to seek a general theoretical answer to the question why criticality 
appears so common, we note that essentially any system in nature is coupled to the 
environmental surroundings and consider information transfer
between the system and the environment. 
%
The mathematical formulation of information was given in the context of 
the communication theory whereas entropy is identified with a measure of 
missing information~\cite{SW}. 
The importance of such information transfer together with the role of entropy 
has been addressed in biological evolution based on random mutation and 
natural selection~\cite{mychoi}: In general, every species tends to minimize its entropy, 
or in other words, attempts to get information from the environment. 
%
Here we propose that this information-transfer dynamics
may serve as a generic and universal mechanism for dynamic scale-invariant behaviors
observed in a vareity of systems, including physical as well as biological 
systems.  Our proposal is then supported by analytical and numerical 
studies of the Ising models as well as biological evolution models.

We consider a system consisting of $N$ elements, the configuration of the $i$th 
of which is represented by $\sigma_i$ in general, together with its environment. 
The total entropy $S_t$ of the whole is given by the sum of the entropy $S$ 
of the system we are interested in and the entropy $S_0$ of the environment: 
$S_t = S + S_0$.
The equilibrium probability $P({\vec \sigma})$ for the system to be in 
configuration ${\vec \sigma} \equiv \{\sigma_1 , \sigma_2 , \ldots, \sigma_N \}$ 
should be proportional to the number $\Omega_0 \, (=e^{S_0}$ with the Boltzmann 
constant $k_B \equiv 1$) of the corresponding accessible states for 
the environment~\cite{mychoi}:
\begin{equation}  \label{eq:Px}
P({\vec \sigma}) \propto e^{S_0} = e^{S_t - S} \equiv C e^{-S}. 
\end{equation} 
When the information transfer between the system and the environment takes place 
in a reversible way, the total entropy does not change in time, keeping $C$ constant. 
(The case of irreversible information transfer will be discussed later.) 

The time evolution of the system is then governed by the master equation for 
the probability $P({\vec \sigma}; t)$ of the configuration ${\vec \sigma}$
at time $t$:
\begin{equation}
\label{eq:master}
\frac{dP(\vec{\sigma}; t)}{dt} =
\sum_{\vec{\sigma}'}[w(\vec{\sigma}' \rightarrow \vec{\sigma}) P(\vec{\sigma}'; t) 
 - w(\vec{\sigma} \rightarrow \vec{\sigma}') P(\vec{\sigma}; t)].
\end{equation}
The transition rate $w({\vec \sigma} \rightarrow {\vec \sigma}^\prime)$ from 
configuration ${\vec \sigma}$ to 
${\vec \sigma}^\prime \equiv \{\sigma'_1 , \sigma'_2 , \ldots, \sigma'_N \}$ 
satisfies the detailed balance condition
\begin{equation} \label{eq:W}
\frac{w({\vec\sigma} \rightarrow {\vec\sigma}^\prime )}
     {w({\vec\sigma}^\prime \rightarrow {\vec\sigma} )}
   = e^{-\{S[E({\vec\sigma}^\prime)] - S[E({\vec\sigma})]\}},
\end{equation}
which ensures that $P({\vec \sigma})$ in Eq.~(\ref{eq:Px}) 
is reached in equilibrium. 
Note also that the entropy in general depends on the configuration through the energy $E$
(or the fitness function in the evolution model~\cite{mychoi}). 
As a simple choice satisfying the detailed balance condition~(\ref{eq:W}), 
one may take the transition rate in the form~\cite{w}
\begin{equation} 
\label{eq:W2}
\tau_0 w(\vec{\sigma} \rightarrow \vec{\sigma}') = \frac{1}{1 + e^{\Delta S}}
= \frac{1}{2}\left(1-\tanh \frac{\Delta S}{2}\right),
\end{equation} 
where $\Delta S \equiv S[E({\vec\sigma}')] - S[E({\vec\sigma})]$ is the entropy change
and $\tau_0$ sets the time scale. 

It is straightforward to obtain the solution of the above master equation 
for the globally-coupled Ising model with the Hamiltonian 
\begin{equation} \label{eq:H}
{\cal H} = -\frac{J}{2N}\sum_{i \neq j}^N \sigma_i \sigma_j,
\end{equation}
where $\sigma_i = \pm 1$ now represents the Ising spin at site $i$
and the coupling strength $J$ is set equal to unity henceforth. 
We need to consider the flip of a single spin, 
i.e, $\sigma'_k = -\sigma_k$ and $\sigma'_i = \sigma_i$ for $i \neq k$,
under which the entropy change $\Delta S$ in Eq.~(\ref{eq:W2})
is obtained as follows: 
The number of configurations in which there are $N_+$ up-spins (having 
the value $+1$) and $N_-$ down-spins (the value $-1$) is given by 
$\Omega = N!(N_+! N_-!)^{-1}$, which gives the entropy $S(M) = \ln\Omega$ 
as a function of the total spin $M \equiv N_+ -N_-$~\cite{mag}.
Under the single-spin flip, the total spin $M$ changes
to $M' = M - 2\sigma_k$, and the corresponding entropy change $\Delta S$, 
combined with Eq.~(\ref{eq:W2}), leads to the transition rate
\begin{equation} 
w(\sigma_k \rightarrow -\sigma_k ) = \frac{N-\sigma_k M+2}{2\tau_0 (N+1)}.
\end{equation} 

Multiplying both sides of Eq.~(\ref{eq:master}) by $\sigma_i$
and taking the summation over all configurations ${\vec\sigma}$,
we obtain
\begin{equation} 
\label{eq:mt1}
\tau_0 \frac{d m}{dt} = -\frac{2}{N+1} m,
\end{equation} 
where $\langle \sigma_i (t)\rangle \equiv \sum_{\vec\sigma} \sigma_i P({\vec\sigma}; t)$,
being independent of $i$, has been recognized as the magnetization $m$
per site.  It is thus concluded that the relaxation time $\tau$ defined by 
$m(t) \sim \exp(-t/\tau)$ scales with the system size as 
\begin{equation} 
\label{eq:tau}
\tau \sim N
\end{equation} 
for large $N$.
Similarly, the correlation function $\langle \sigma_i (t) \sigma_j (0) \rangle$ 
can be computed by multiplying Eq.~(\ref{eq:master}) by $\sigma_i \sigma'_j$
with the initial configuration ${\vec\sigma}'$ imposed at time $t=0$. 
This leads to the same relaxation $\langle M(t) M(0) \rangle \sim \exp(-t/\tau)$ 
with $M \equiv \sum_i \sigma_i$. 
The energy relaxation also obtains the same form, with half the relaxation time. 
Physically, $\tau$ describes the time scale 
required to overcome the entropic barrier~\cite{entropicbarrier}
separating the two states $m = 1$ and $m = -1$. 
The above scaling behavior has a profound implication that the information-transfer 
dynamics does not possess a characteristic time scale in the thermodynamic limit, 
which may be interpreted as that the system becomes critical in equilibrium.
Indeed the power spectrum ${\cal P}(\omega) \sim (\omega^2 +\tau^{-2})^{-1}$
at frequency $\omega$, easily obtained from the exponential relaxation,
reduces to the pure power-law form ${\cal P}(\omega) \sim \omega^{-2}$ 
as $\tau$ grows large with $N$; 
such $1/\omega^2$ power spectra have been observed 
in a number of {\em specific} systems studied mostly as models for SOC~\cite{ref:f2}. 
The linear dependence on the system size $N$ also implies that the dynamic 
critical exponent $z$ defined by $\tau \sim L^z$ with the linear size $L$ 
has the value four since the upper critical dimension of the Ising model is 
well known to be four. 
Equation~(\ref{eq:mt1}), yielding $dm/dt = 0$ in the thermodynamic limit, 
corresponds to conserved dynamics, for which the mean-field value $z=4$ 
is expected~\cite{dynamic}. 

From the viewpoint of numerical simulations, it is of interest 
that the detailed balance condition in Eq.~(\ref{eq:W}) precisely corresponds
to the entropic sampling algorithm~\cite{mychoi,jlee}. 
Namely, dynamic simulations based on the entropic sampling algorithm naturally describe 
the time evolution of the system in which the information transfer to and from 
the environment is reversible.  In that algorithm~\cite{jlee}, 
one initially starts from $S(E) = 0$ for all values of the energy $E$, 
and obtains the histogram $H(E)$ for several Monte Carlo sweeps (MCS), which
is then used to estimate a new value of $S(E)$:
\begin{equation}
S(E) = \left\{
\begin{array}{ll}
S(E) & \mbox{for $H(E) = 0$, } \\
S(E) + \ln H(E)  & \mbox{otherwise.} \\
\end{array}
\right.
\end{equation}
As the above procedure proceeds, $S(E)$ approaches the true entropy up to
an additive constant; this is independent of $E$ and may thus be subtracted, 
according to the condition that the minimum of $S(E)$ vanishes. 
Once the correct entropy is obtained, the time evolution under the
entropic sampling should satisfy the detailed balance condition in Eq.~(\ref{eq:W}) 
and we take the following procedure: 
(i) Generate configuration ${\vec \sigma}^\prime$ which differs from ${\vec \sigma}$ 
only at a single site, say, $k$. 
(ii) Compute the entropy change 
$\Delta S \equiv S[E({\vec\sigma}^\prime)]-S[E({\vec\sigma})]$. 
(iii) If $\Delta S \leq 0$, accept the try and change ${\vec\sigma}$ to 
${\vec\sigma}^\prime$; otherwise, accept the try with the probability $e^{-\Delta S}$. 
One sweep of the above procedure for all $N$ constituents in the system corresponds
to one time unit. 

\begin{figure}
\centering{\resizebox*{!}{12cm}{\includegraphics{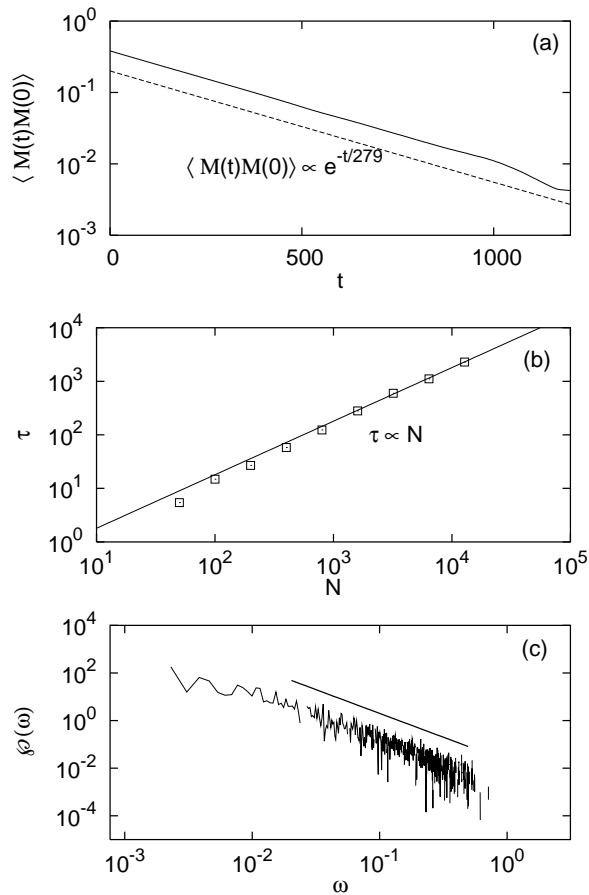}}}
\caption{Entropic sampling dynamics in the globally-coupled Ising model, 
with quantities expressed in arbitrary units. 
(a) Correlation function $\langle M(t) M(0) \rangle$ 
in the system of size $N=1600$.  The dashed line represents the exponential function
$\exp(-t/\tau)$ with the relaxation time $\tau =279$. 
(b) Relaxation time $\tau$ versus system size $N$ 
in the log-log scale, displaying the relation $\tau \sim N$ 
represented by the solid line. 
(c) Power spectrum ${\cal P}(\omega)$ as a function of frequency $\omega$, 
obtained from the Fourier transform of the correlation function 
in the system of size $N=1600$, 
in comparison with the power-law form ${\cal P}(\omega) \sim \omega^{-2}$ 
represented by the straight line.
}
\label{fig:global}
\end{figure}
In the above manner we perform entropic sampling dynamic simulations on the
globally-coupled Ising model, and compute the relaxation time $\tau$ from the correlation
function $\langle M(t) M(0) \rangle$. 
In practice, assuming ergodicity and time-translation symmetry in equilibrium, 
we take the time average over $t^\prime$ in the expression 
$\langle M(t{+}t^\prime) M(t^\prime)\rangle_{t^\prime}$,
and display the results in Fig.~\ref{fig:global}. 
The correlation function, shown in (a), indeed follows the exponential relaxation
$\langle M(t) M(0) \rangle \sim \exp(-t/\tau)$. 
The relaxation time $\tau$ is obtained from the least-square-fit and plotted 
as a function of the system size $N$ in (b), 
which confirms the analytical result $\tau \sim N$.  
The power spectrum ${\cal P}(\omega)$ is also computed as a function of the 
frequency $\omega$
and shown in (c). 
We have considered various sizes and observed that 
the frequency region described by the power-law ${\cal P}(\omega) \sim \omega^{-2}$ 
indeed extends with the size $N$,
which manifests the emerging critical behavior in the thermodynamic limit. 
We have also computed the power spectrum of the energy and found the same behavior, 
in agreement with the analytical result. 

\begin{figure}
\centering{\resizebox*{!}{8cm}{\includegraphics{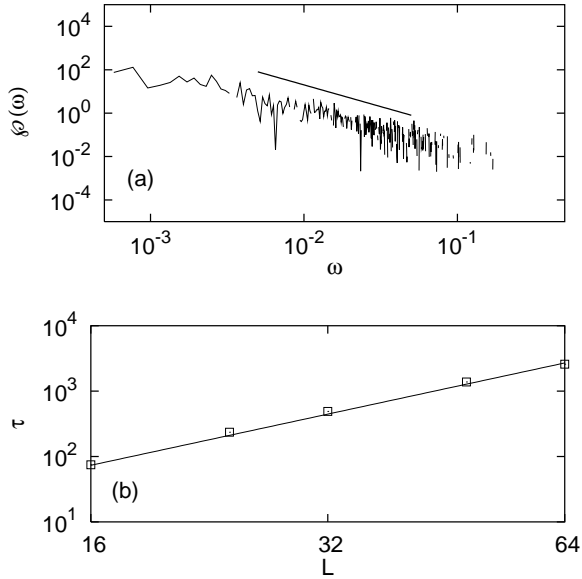}}}
\caption{The two-dimensional Ising model, with quantities expressed in 
arbitrary units. 
(a) Power spectrum ${\cal P}(\omega)$ of the magnetization in the system of size $L=32$,
together with the power-law form $\omega^{-2}$ represented by the straight line.
(b) Relaxation time $\tau$
versus the system size $L$: Observed is $\tau \sim L^{2.6}$, represented by the solid line. 
}      
\label{fig:2d}
\end{figure}
It is necessary to check whether the scale-invariant criticality, 
observed above in the globally-coupled system, 
is merely an artifact of the mean-field model. 
For this, we study the two-dimensional Ising model on $L\times L$ square lattices, 
which contains only local interactions and thus provides a more realistic system. 
The price of this is lack of analytic tractability, and accordingly we
rely on numerical methods.  We thus follow the entropic sampling algorithm
to compute the entropy $S(E)$ first and also the specific heat 
from $S(E)$.  The latter is found to display a peak near temperature $T = 2.27$ 
in the large system of $L=64$, the comparison of which with the precise critical
temperature $T_c = 2.2692...$ demonstrates the reliability of the obtained data. 
Via the same procedure, we measure the magnetization and the energy, as well as 
their power spectra [see Fig.~\ref{fig:2d}(a)]. 
The relaxation time $\tau$, estimated from the relaxation data, 
is shown in Fig.~\ref{fig:2d}(b). 
Observed is the algebraic increase with the size:
$\tau \sim L^z$ with $z \approx 2.6$. 
The nontrivial value of the dynamic exponent $z$ suggests 
that the observed criticality is a genuine emergent property resulting from 
cooperative phenomena 
and may not be explained by simple rescaling of simulation time steps. 

\begin{figure}
\centering{\resizebox*{!}{8cm}{\includegraphics{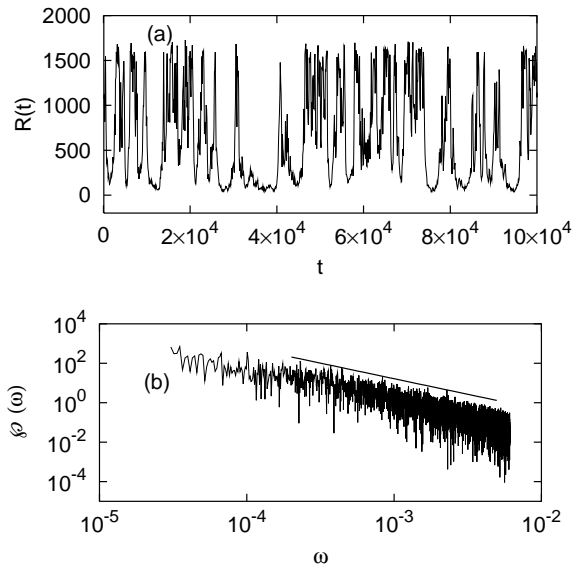}}}
\caption{The biological evolution model in Ref.~\cite{mychoi}, with quantities 
expressed in arbitrary units. 
(a) Time evolution of the mutation rate $R(t)$ with the data compiled in time steps
of $\Delta t =10$
after the initial transient time $2.5\times 10^6$. 
(b) Power spectrum ${\cal P}(\omega)$ of $R(t)$, 
exhibiting characteristic power-law behavior ${\cal P}(\omega) \sim \omega^{-1.5}$ 
represented by the solid line. 
}
\label{fig:evolve}
\end{figure}

In order to establish the universality of these findings further, 
we also investigate one of biological evolution models, where
random mutation and natural selection leads naturally to entropic sampling
dynamics~\cite{mychoi}. 
Here we focus on the dynamic behavior of the mutation rate, i.e., 
how the number of mutations during given time interval varies with time. 
Figure~\ref{fig:evolve} displays the typical behavior in the one-dimensional 
system of $256$ species with the Bak-Sneppen type fitness function~\cite{mychoi,bs}:
(a) time series of the mutation rate $R(t)$ compiled in time steps of $10$ MCS and 
(b) the corresponding power spectrum ${\cal P}(\omega)$. 
(We have varied the time step and considered various forms of the fitness function 
also in higher dimensions, only to find no qualitative difference.) 
Again confirmed is the SOC behavior, characterized by the absence
of the characteristic time scale and ${\cal P}(\omega) \sim \omega^{-\alpha}$
with $\alpha = 1.5 \pm 0.1$. 
It is pleasing to note that mutations correspond to extinction events of species,
for which fossil data indeed yield similar behavior of the power spectrum~\cite{bio}. 

From these studies of several model systems, we expect the ubiquity of emerging 
dynamic criticality regardless of the details of the system, 
as long as the dynamics is directed by information transfer to and from the environment. 
In this respect, the information-transfer dynamics as a general mechanism for 
emerging criticality appears to provide a theoretical explanation why scale invariance 
is observed so common in nature.

Finally, we consider the situation in which information transfer is not reversible. 
In this irreversible case, the total entropy does not remain constant but increases as 
the energy (or the fitness) of the system is reduced (or raised) via information transfer. 
We expand the total entropy around some reference energy (or fitness) $E_0$:
\begin{equation} 
S_t(E) \approx 
S_t(E_0) + (E-E_0)\left.\frac{\partial S_t}{\partial E}\right|_{E_0},
\end{equation} 
and obtain $C= e^{-\beta E}$ in Eq.~(\ref{eq:Px}), i.e.,
the probability depending on the energy as well as the entropy
\begin{equation} 
\label{eq:Px:irrerversible}
P({\vec \sigma}) \propto e^{- S - \beta E},
\end{equation}
where $\beta \equiv -(\partial S_t/\partial E)_{E_0}$ controls
the amount of irreversible information transfer.  In the absence of the energy term 
($\beta = 0$), information transfer is fully reversible and the entropic sampling 
is recovered; conversely, the absence of the entropy term leads to 
the standard Metropolis Monte Carlo algorithm. 
The transition rate corresponding to the equilibrium probability in 
Eq. (\ref{eq:Px:irrerversible}) may be taken as
\begin{equation} \label{other}
w({\vec\sigma} \rightarrow {\vec\sigma}' ) =  
\frac{1}{2\tau_0}\left[1-\tanh\frac{1}{2}(\Delta S+\beta\Delta E)\right]. 
\end{equation}
In the globally-coupled Ising model,
this leads to the evolution equation for the magnetization (in the thermodynamic limit):
\begin{equation} 
\label{eq:mt2}
\frac{d m}{dt} = -m + \frac{ m + \tanh\beta J m }{ 1 + m \tanh\beta J m},
\end{equation} 
which in the reversible limit ($\beta \rightarrow 0$) reduces to Eq.~(\ref{eq:mt1}) with 
$N \rightarrow \infty$.  
When $\beta J$ is sufficiently small, a simple expansion gives
$dm/dt = \beta J( m - m^3)$, which is easily solved to yield the
relaxation time $\tau = (2\beta J)^{-1}$.
Strictly speaking, as soon as $\beta$ is raised from zero, 
the relaxation time becomes finite, destroying scale invariance. 
However, when the irreversibility is weak, $\beta$ may be so small that the relaxation 
time can still be much longer than the observation time.  
Then deviations from the scale-invariant behavior may not be observed in the practical 
sense~\cite{finite}.

In summary, we have investigated dynamics based on information transfer,
revealing the emergence of dynamic critical behavior.
This is implicative to a wide range of phenomena in biology and sociology
~\cite{babinec}
as well as in physics, and expected to be of broad applicability.

We thank J. Lee for helpful discussions
and acknowledge partial support from the KOSEF
Grant Nos.\ R01-2002-000-00285-0 and R14-2002-062-01000-0, from 
the BK21 Program, and from Ajou University. 


\begin{thebibliography}{99}
\bibitem{w} Other forms satisfying the detailed balance condition may be chosen 
without altering the subsequent analysis.
\bibitem{mag} Note that the energy of the globally-coupled model is uniquely specified 
by the magnetization: $E=-M^2 /2N$.
\bibitem{finite} Note the similarity to the finite-size effects:  Any system in nature 
as well as in numerical study is in principle finite, and accordingly the relaxation time 
is finite, albeit long, even in the reversible case.  Still scale invariance is observed 
in practice, as demonstrated in Figs. \ref{fig:global} - \ref{fig:evolve}.
\end{thebibliography}
\end{document}